\documentstyle[a4wide,12pt,epsf]{article}
\begin{document}
%\begin{frontmatter}
\begin{flushright}
\baselineskip=12pt
{SUSX-TH-98-002}\\
{hep-th/9804135}\\
{March 1998}
\end{flushright}
\begin{center}
{\LARGE \bf The effect of Wilson line moduli on  
CP-violation by soft supersymmetry breaking terms. }
\vglue 0.45 cm
%\author[SUSSEX]{D. Bailin\thanksref{gramma}}
%\author[RHBNC]{G.V. Kraniotis\thanksref{episto}}
%\author[RHBNC]{A. Love}
%\address[SUSSEX]{Centre for Theoretical Physics,
%University of Sussex,Brighton BN1 9QJ, U.K.}
%\address[RHBNC]{Department of Physics, 
%Royal Holloway and Bedford New College,University of London,Egham, 
%Surrey TW20-0EX, U.K.}
%\thanks[gramma]{D.Bailin@sussex.ac.uk}
%\thanks[episto]{G.Kraniotis@rhbnc.ac.uk}
{D. BAILIN$^{\clubsuit}$ \footnote
{D.Bailin@sussex.ac.uk}, G.V. KRANIOTIS$^{\spadesuit}$ \footnote
{G.Kraniotis@rhbnc.ac.uk} and A. LOVE$^{\spadesuit}$ \\}
{$\clubsuit$ \it Centre for Theoretical Physics, \\}
{\it University of Sussex,\\}
{\it Brighton BN1 9QJ, U.K. \\}
{$\spadesuit$ \it Department of Physics, \\} 
{\it Royal Holloway and Bedford New College. \\}
{\it University of London, Egham, \\}
{\it Surrey TW20-0EX, U.K. \\}
\baselineskip=12pt

\vglue 0.25cm
%ABSTRACT
\end{center}
{\rightskip=3pc
\leftskip=3pc
\noindent
\baselineskip=20pt
\begin{abstract}
The $CP$-violating phases in the soft supersymmetry-breaking 
sector in orbifold compactifications with a continuous Wilson 
line are investigated. 
In this case the modular symmetry is the Siegel
modular  group
$Sp(4,Z)$ of genus two.
In particular, we study the case that the hidden sector 
non-perturbative superpotential is determined by the Igusa 
cusp form ${\cal C}_{12}$ of modular weight 12.
The effect of large non-perturbative corrections to 
the dilaton K$\rm{\ddot a}$hler potential on the resulting 
$CP$-violating phases is also investigated. 
\end{abstract}
%\end{frontmatter}

\newpage
Duality symmetries in string theory have deep implications for the
moduli dependence of the effective action of the resulting 
supergravity theory. In particular the
moduli dependence of threshold corrections to the 
gauge couplings entails various automorphic forms of the corresponding 
duality group. In $N=1$ effective supergravities from string theory 
the moduli dependence of the K${\rm {\ddot a}}$hler 
potential and of the non-perturbative 
superpotential $W_{np}$ is also very constrained. In particular 
$W_{np}$ has to transform as a {\it modular form} under duality 
transformations in order that the gravitino mass is an invariant of the 
modular group \cite{SERGIO}. 
The transformation of $W_{np}$ as a modular form 
as described  above has been recently
noted in  the strong  
coupling case by Donagi et al \cite{WITTEN} 
who showed that there exist compactifications  
in $F$ and $M$-theory, for instance 
on the Calabi-Yau fourfold $X$ with configuration matrix, 
$$\left[\begin{array}{c}
2 \\ 1 \\ 1 \\ 2\end{array}\right|
\left|\begin{array}{cc}
3 & 0 \\
1 & 1 \\
0 & 2 \\
0 & 3\end{array}\right]$$
in which the emerging $N=1$
$W_{np}$ has modular properties. In fact it is an $E_8$ theta function.

The main modular forms that have appeared in the effective action, 
besides the 
$E_8$ theta function modular form in $W_{np}$, are 
the Dedekind eta function $\eta(T)$ \cite{DKL} 
and the absolute modular invariant 
$j(T)$ (in gauge group independent threshold
corrections) \cite{KRP}. Automorphic functions 
of the Siegel modular group for genus-2 have appeared in threshold 
corrections in 
$N=2$ compactifications \cite{KAW} and have also arisen in  
the counting of microstates in certain 
stringy black hole systems \cite{RDH}. The 
Igusa cusp form of weight 10, ${\cal C}_{10}$,
is the particular modular form involved.
Siegel modular forms have also appeared in the effective action 
in the study of string duals of $N=2,d=4$  heterotic compactifications 
on $K3\times T^2$ and type $IIA$ compactifications on suitably 
chosen Calabi-Yau threefolds \cite{LUST}.
The Siegel modular group is intimately connected to the symplectic geometry.
%%The special K${\rm {\ddot a}}$hler 
%%geometry is the geometry of the moduli space of 
%%Calabi Yau compactifications. 
In addition, and more specifically, Mayr and Stieberger \cite{MS} 
and Nilles and Stieberger \cite{NST} have proposed the use of genus-2
Siegel modular forms in the gauge kinetic function and 
in threshold corrections
in $N=1$ orbifold compactifications.
All the above interesting results strongly
motivate  the study of the effective string 
supergravity in which   $W_{np}$ transforms as a Siegel modular 
form.

We have previously studied the implications
of $PSL(2,Z)$ and $\Gamma^{0}(3)$ 
duality-invariant effective actions for the $CP$-structure of
string theory \cite{BKL}. We showed that the $CP$-violating phases 
in the soft supersymmetry breaking terms are related to the properties 
of the modular functions involved in $W_{np}$. 
Specifically we showed that zero or very 
small ($\leq 10^{-4}$)  $CP$-violating phases from the 
soft supersymmetry-breaking $A$ and $B$ terms 
arise for minima of the 
non-perturbative effective potential at complex values of the moduli 
{\em on } the boundary of the 
standard ``fundamental domain'' of the modular group;
in principle minima might also arise at interior points of the 
fundamental domain (in which case larger 
phases do arise), but it appears that this is only possible for 
unphysical values of the dilaton kinetic terms. Values of the 
moduli at the minimum of the effective potential on the unit circle 
and in the interior of the standard fundamental domain of $PSL(2,Z)$ 
were obtained in the presence of the absolute modular invariant $j(T)$
in $W_{np}$.
In this paper we 
extend our previous results by investigating the case in which 
a continuous Wilson line $B$ is also present in the effective 
supergravity besides the $T$-  and $U$-moduli \cite{MS,NST}.
In this case the 
modular symmetry of the effective supergravity  
is the genus two Siegel modular group 
$Sp(4,Z)$. In particular as suggested by Mayr and Stieberger \cite{MS}, 
 we study the case in which 
the Igusa cusp form ${\cal C}_{12}$ of weight 12 appears in the hidden 
sector non-perturbative superpotential. 
${\cal C}_{12}$  is the generalization of the Dedekind eta function
$\eta(T)$ which is the modular form 
present in the hidden sector $W_{np}$ with a  $PSL(2,Z)$ modular symmetry.
To estimate the size of the $CP$-violating phases one has to minimize the
effective potential $V_{eff}$ with respect to all of the moduli.

%The non-perturbative superpotential resulting from a gaugino 
%condensation in the hidden sector which reflects the Siegel's 
%symplectic geometry for genus 2 is given by:
%\begin{equation}
%W_{np}={\mit \Omega(S)}{{\cal C}_{12}(\Omega)}^{-1/12}
%\end{equation}

For the $Z_8$ orbifold 
considered by Mayr and Stieberger in the presence of Wilson line moduli $B,C$, 
besides the usual $T,U$ moduli (in the first complex plane)
the perturbative K$\rm {\ddot a}$hler potential correct to quadratic 
order in matter fields is given by
\begin{eqnarray}
K&=&-\log(y)-\sum_{i=2,3}\log(T_i+\bar{T_i})-\log D+
\sum_{\alpha} D^{p_{\alpha}}\prod_{i=2,3} (T_i+\bar{T}_i)^{n_{\alpha}^i}
\Phi_{\alpha}\Phi_{\bar{\alpha}} \nonumber \\
&+& D^{-1} \phi_1 \phi_2 +{\rm h.c. }
\label{one}
\end{eqnarray}
where 
\begin{equation}
D=(T+\bar{T})(U+\bar{U})-(B+\bar{C})(C+\bar{B})
\end{equation}
\begin{equation}
y=S+\bar{S}-\sum_{i=2,3}\delta_i \log(T_i+\bar{T}_i)
\end{equation}
where  $S$ is the dilaton, $T_i$, $i=2,3$ are the  $N=1$ moduli,
and $\delta_i$ are the Green-Schwarz anomally cancellation coefficients.
$p_{\alpha},n_{\alpha}^i$ are the modular weights of the 
matter fields $\Phi_{\alpha}$. In the special case of the untwisted 
matter fields $\phi_1,\phi_2$ associated with the first complex plane 
$p_1=p_2=-1$ and $n_{1}^{i}=n_{2}^{i}=0$.

In the case that large non-perturbative corrections 
to the dilaton K$\rm {\ddot a}$hler potential
are responsible for the stabilization of the dilaton field,
the K$\rm {\ddot a}$hler potential is more generally given 
by:
\begin{eqnarray}
K&=&P(y)-\log(T_i+\bar{T_i})-\log D+
\sum_{\alpha} D^{p_{\alpha}}\prod_{i=2,3} (T_i+\bar{T}_i)^{n_{\alpha}^i}
\Phi_{\alpha}\Phi_{\bar{\alpha}} \nonumber \\
&+&D^{-1} \phi_1 \phi_2 +{\rm h.c. }
\end{eqnarray}
where  $P(y)$ is a function to be determined by 
stringy non-perturbative effects.
In that case, we shall treat $\frac{dP}{dy}$ and 
$\frac{d^2 P}{dy^2}$, which we shall see, occur in the 
effective potential and the soft supersymmetry-breaking terms, 
as free parameters. We require that  $\frac{d^2 P}{dy^2}>0$ so that 
the dilaton kinetic terms have the correct sign.

As observed by Mayr and Stieberger \cite{MS}, the construction of 
a superpotential involving Wilson lines for the $Z_8$ orbifold and 
having the correct $Sp(4,Z)$ modular covariance, can only be 
achieved in the case that $B=C$. Then $W_{np}$ arising from hidden 
sector condensation is given by \footnote{The Igusa cusp form
${\cal C}_{12}(\Omega)$
\cite{IGUSA}, can be expressed as a certain combination 
of genus-2 theta functions with characteristics, 
${\cal C}_{12}(\Omega)=(3\times 2^{17})^{-1}\sum(\Theta_{m1}
\Theta_{m2}\cdots\Theta_{m6})^{4}$. 
The summation is extended over the fifteen compliments of the 
so called G$\rm{\ddot{o}}$bel quadruples. 
A G$\rm{\ddot{o}}$bel quadruple consists of four distinct 
even characteristics which form a syzygous sequence.}
\begin{equation}
W_{np}=F(S){{\cal C}_{12}(\Omega)}^{-1/12}
\label{dyna}
\end{equation}
where $F(S)$ gives the, in general unknown, dependence upon the 
dilaton, and 
$$\Omega=\Bigl(\begin{array}{cc}
T & B \\
B & U
\end{array}\Bigr)$$
However, in the case of a single gaugino condensate $F(S)\propto
e^{\frac{24\pi^2 S}{b}}$ is known.
The effective potential  is then given by:
\begin{eqnarray}
e^{-P(y)}\prod_{i=2,3}(T_i+\bar{T}_i)V_{eff}&=&D^{-1} |W_{np}|^2 
\Biggl\{(\frac{d^2 P}{d y^2})^{-1} \Bigl|\frac{dP}{dy} 
+\frac{\partial{\log W_{np}}}{\partial S}
\Bigr|^2-2 \nonumber \\
%%&+& \sum_{i=2,3} (1+\delta_{i} \frac{dP}{dy})^{-1} \Bigl|
%%(T_i+\bar{T}_i)\frac{\partial{\log W_{np}}}{\partial T_i}-1 
%%+\delta_i \frac{\partial{\log W_{np}}}{\partial S}\Bigr|^2  \nonumber \\
&+&\Bigl|1-(T+\bar{T})\frac{\partial {\log W_{np}}}{\partial T} \nonumber \\
&-&(U+\bar{U})\frac{\partial {\log W_{np}}}{\partial U}-
(B+\bar{B})\frac{\partial {\log W_{np}}}{\partial B}\Bigr|^2 \nonumber \\
&+& D \Bigl(
\frac{1}{2}|\frac{\partial {\log W_{np}}}{\partial B}|^2-
(\frac{\partial {\log W_{np}}}{\partial T} \frac{\partial 
{\log \bar{W}_{np}}}{\partial{\bar{U}}}+{\rm h.c.})\Bigr)\Biggl\} \nonumber \\
\label{moddyn}
\end{eqnarray}
The $N=1$ moduli $T_2,T_3$, do not contribute 
to the right hand side of (\ref{moddyn})
and we have also set $\delta_{GS}^{i}=b/3$ as is appropriate for 
a pure gauge hidden sector.

%In the limit $B=C$ if 
%$\tilde{W_{np}}(B)\equiv W_{np}(B,C=B)$, then
%\begin{equation}
%\frac{\partial{\log W_{np}}}{\partial B}=
%\frac{\partial {\log W_{np}}}{\partial C}=\frac{1}{2}
%\frac{\partial {\log \tilde{W}_{np}}}{\partial B}
%\end{equation}

%In the moduli dominated limit in which the dilaton $F_s$ term 
%does not contribute to the supersymmetry breaking $V_{eff}$
%reduces to 
%\begin{eqnarray}
%(S+\bar{S})\prod_{i=2,3}(T_i+\bar{T_i})\;D \;V_{eff}&=&|W_{np}|^2\Biggl\{-2 
%\nonumber \\
%&+&|1-(T+\bar{T})\frac{\partial \log W_{np}}{\partial T}-
%(U+\bar{U})\frac{\partial \log W_{np}}{\partial U} \nonumber \\
%&-&(B+\bar{B})\frac{\partial \log W_{np}}{\partial B}|^2 \nonumber \\
%&+&D\Bigl(\frac{1}{2}\Bigl|\frac{\partial \log W_{np}}{
%\partial B}\Bigr|^2-(\frac{\partial \log W_{np}}{\partial T}
%\frac{\partial \log \bar{W_{np}}}{\partial \bar{U}}+\rm{h.c})
%\Bigr)\Biggr\} \nonumber \\
%\end{eqnarray}

We now minimize  $V_{eff}$,  and  calculate 
the  soft 
supersymmetry-breaking $A$ and 
$B$ terms and study the $CP$-properties of 
the theory with a Wilson line present.
The soft trilinear $A$- term 
associated with the term $h_{\alpha\beta\gamma}\Phi_{\alpha}
\Phi_{\beta}\Phi_{\gamma}$ of the perturbative superpotential is 
 given by 
\begin{eqnarray}
-m_{3/2}^{-1} A_{\alpha\beta\gamma}&=&(\frac{d^2 P}{d y^2})^{-1} 
\Bigl(\frac{dP}{dy}+\frac{24 \pi^2}{b}\Bigr)
\frac{dP}{dy} \nonumber \\
%&-&\sum_{i=2,3}(1+\delta_i\frac{dP}{dy})^{-1}(\delta_i
%\frac{24 \pi^2}{b}-1)\Bigl(1+n_{\alpha}^{i}+n_{\beta}^{i}+
%n_{\gamma}^{i}-(T_i+\bar{T}_i)\frac{\partial 
%\log{h_{\alpha\beta\gamma}}}{\partial T_i}\Bigr) \nonumber \\
&-&\Bigl((T+\bar{T})\frac{\partial {\log \bar{W}_{np}}}{\partial 
\bar{T}}+(U+\bar{U})\frac{\partial{\log \bar{W}_{np}}}{\partial \bar{U}}+
(B+\bar{B})\frac{\partial {\log \bar{W}_{np}}}{\partial \bar{B}} -2\Bigr) 
\nonumber \\
&\times&\Bigl(1+p_{\alpha}+p_{\beta}+p_{\gamma}-
(T+\bar{T})\frac{\partial \log h_{\alpha\beta\gamma}}{\partial T}-
(U+\bar{U})\frac{\partial \log h_{\alpha\beta\gamma}}{\partial U} \nonumber \\
&-&(B+\bar{B})\frac{\partial 
\log h_{\alpha\beta\gamma}}{\partial B} \Bigr)\nonumber \\
&+&\Bigl((T+\bar{T})-D\frac{\partial {\log \bar{W}_{np}}}{\partial
\bar{U}}\Bigr)\frac{\partial \log h_{\alpha\beta\gamma}}{
\partial T}+\Bigl((U+\bar{U})-D \frac{\partial \log \bar{W}_{np}}{
\partial \bar{T}}\Bigr)\frac{\partial \log h_{\alpha\beta\gamma}}{
\partial U} \nonumber \\
&+&\Bigl((\bar{B}+B)+\frac{1}{2}D \frac{\partial \log \bar{W}_{np}}
{\partial \bar{B}}\Bigr)\frac{\partial \log h_{\alpha\beta\gamma}}
{\partial B}\nonumber \\
\end{eqnarray}

It is rather natural to identify $\phi_1,\phi_2$ with Higgs scalars 
because the mixing term in the K${\rm {\ddot a}}$hler potential $K$ 
provides an effective $\mu$ term.
The soft $B$ term associated with this mixing term 
in $K$, and a possible bilinear term 
$\mu_W $ in  the superpotential  
is given by
\begin{eqnarray}
m_{3/2}^{-1}\mu_{eff}B&=&2 D^{-1} W_{np}-\mu_W \nonumber \\
&+& (\frac{d^2 P}{dy^2})^{-1} \Bigl(\frac{dP}{dy}+
\frac{24 \pi^2}{b}\Bigr) \Bigl(\frac{d P}{d y}+
\frac{\partial \log \mu_W}{\partial S}\Bigr) \mu_W \nonumber \\
%%&-&\sum_{i=2,3}(1+\delta_i\frac{dP}{dy})^{-1}(\delta_i 
%%\frac{24 \pi^2}{b}-1)\Bigl(1-(T_i+\bar{T}_i)
%%\frac{\partial \log \bar{W}_{np}}{\partial \bar{T_i}}-
%%\delta_i\frac{24 \pi^2}{b}\Bigr) \nonumber \\
%%&\times&\Bigl(1+n_1^i+n_2^i-(T_i+\bar{T}_i)
%%\frac{\partial \log \mu_W}{\partial T_i}-\delta_i
%%\frac{\partial \log \mu_W}{\partial S}\Bigr)\mu_W \nonumber \\
&+&\Bigl((T+\bar{T})\frac{\partial \log \bar{W}_{np}}{\partial 
\bar{T}}+(U+\bar{U})\frac{\partial \log \bar{W}_{np}}{\partial
\bar{U}}+(B+\bar{B})\frac{\partial \log \bar{W}_{np}}{\partial
\bar{B}}-2\Bigr) \nonumber \\
&\times&[1+\Bigl((T+\bar{T})\frac{\partial}{\partial T}+
(U+\bar{U})\frac{\partial}{\partial U}+(B+\bar{B})\frac{
\partial}{\partial B}\Bigr)
\log\mu_W]\mu_W \nonumber \\
&+&\Bigl\{\Bigl((T+\bar{T})-D\frac{\partial \log \bar{W}_{np}}{
\partial \bar U}\Bigr) \frac{\partial \log \mu_W}
{\partial T}+\Bigl((U+\bar{U})-D \frac{\partial \log 
\bar {W}_{np}}{\partial \bar{T}}\Bigr)\frac{\log \mu_W}{\partial U} 
\nonumber \\
&+&\Bigl((\bar{B}+B)+\frac{1}{2} D \frac{\partial \log \bar{W}_{np}}
{\partial \bar{B}}\Bigr)\frac{\partial \log \mu_W}
{\partial B}\Bigr)\Bigr\}\mu_W \nonumber \\
&+&D^{-2} W_{np}\Bigl((T+\bar{T})-D\frac{\partial \log \bar{W}_{np}}
{\partial \bar{U}}\Bigr)\Bigl((U+\bar{U})-
D\frac{\partial \log W_{np}}{\partial T}\Bigr) \nonumber \\
&+&D^{-2} W_{np}\Bigl((U+\bar{U})-D\frac{\partial \log 
\bar{W}_{np}}{\partial \bar{T}}\Bigr)
\Bigl((T+\bar{T})-D\frac{\partial \log W_{np}}{
\partial U}\Bigr) \nonumber \\
&-&2 D^{-2} W_{np}\Bigl((\bar{B}+B)+
\frac{1}{2}D \frac{\partial \log \bar{W}_{np}}
{\partial \bar{B}}\Bigr)\Bigl((B+\bar{B})+
\frac{1}{2}D\frac{\partial \log W_{np}}
{\partial B}\Bigr) \nonumber \\
%&-&D^{-2} W_{np}\Bigl((\bar{B}+B)+
%\frac{1}{2}D\frac{\partial \log \bar{W}_{np}}{\partial \bar{B}}\Bigr)
%\Bigl((B+\bar{B})+\frac{1}{2}D\frac{\partial \log W_{np}}{\partial B}\Bigr) 
%\nonumber \\
&+&D^{-1}W_{np}\Bigl[(T+\bar{T})\frac{\partial \log W_{np}}{
\partial T}+(U+\bar{U})\frac{\partial \log W_{np}}{\partial U}+
(\bar{B}+B)\frac{\partial \log W_{np}}{\partial B} -2\Bigr] \nonumber \\
&+&D^{-1}W_{np}\Bigl[(T+\bar{T})\frac{\partial \log \bar{W}_{np}}{
\partial \bar{T}}+(U+\bar{U})\frac{\partial \log \bar{W}_{np}}{\partial 
\bar{U}}+
(\bar{B}+B)\frac{\partial \log \bar{W}_{np}}{\partial \bar{B}} -2\Bigr] \nonumber \\
&+&D^{-1}W_{np}\times\Biggl\{-2+\Bigl|\frac{dP}{dy}+
\frac{24\pi^2}{b}\Bigr|^2(\frac{d^2 P}{dy^2})^{-1} \nonumber \\
%%&-&\sum_{i\not =T,U}\frac{1}{(1+\delta_i\frac{dP}{dy})}
%%|1-(T_i+\bar{T_i})\frac{\partial\log W_{np}}{\partial T_i}-
%%\delta_i \rho|^2 \nonumber \\
&+&\Bigl|1-(T+\bar{T})\frac{\partial \log W_{np}}{\partial T}-
(U+\bar{U})\frac{\partial \log W_{np}}{\partial U}-
(B+\bar{B})\frac{\partial\log W_{np}}{\partial B}\Bigr|^2 \nonumber \\
&+&D\Biggl(\frac{1}{2}\Bigl|\frac{\partial \log W_{np}}{\partial B}
\Bigr|^2-
\frac{\partial \log W_{np}}{\partial T}\frac{\partial \log 
\bar{W}_{np}}{\partial \bar{U}}-
\frac{\partial \log \bar{W}_{np}}{\partial \bar{T}}
\frac{\partial \log W_{np}}{\partial U}\Biggr)\Biggr\} \nonumber \\
\end{eqnarray}

The effective $\mu$ term is given by
\begin{eqnarray}
\mu_{eff}&=&\frac{|W_{np}|}{W_{np}}e^{\frac{K}{2}} \nonumber \\
&\times&\Biggl\{\mu_W+W_{np}D^{-1} \nonumber \\
&+&W_{np}D^{-1}\Bigl[(T+\bar{T})\frac{\partial \log W_{np}}
{\partial T}+(U+\bar{U})\frac{\partial \log W_{np}}
{\partial U} \nonumber \\
&+&(B+\bar{B})\frac{\partial \log W_{np}}{\partial B}-2\Bigr]\Biggr\}
\nonumber \\
\end{eqnarray}
To go to the low energy supergravity we need to rescale $B$ by 
a factor $\frac{|W_{np}|}{W_{np}}e^{K/2}$ which then cancels when 
dividing by $\mu_{eff}$.

We first minimize $V_{eff}$ in the moduli dominated case, i.e 
$\frac{dP}{dy}+\frac{24\pi^2}{b}=0$. As a first check of our calculation
we find the minimum of $V_{eff}$ for the case the Wilson line is 
turned off, i.e. $B=0$.
Then \begin{equation}
{\cal C}_{12}(0,T,U)=\Delta(T)\Delta(U)
\end{equation}
where 
\begin{equation}
\Delta(T_{i})={\eta(T_i)}^{24}
\end{equation}
This fact has been demonstrated both analytically  
 and verified numerically. The minimum in the 
moduli dominated limit is at $T_{min}=U_{min}\sim 1.2$, in accordance 
with previous results \cite{SPAIN}.
Now we turn on the Wilson line and we obtain the minimum (see fig.1)
\begin{eqnarray}
T_{min}=1.4643126+0.5625414\;i \nonumber \\
B_{min}=0.3347585+0.1300201\;i \nonumber \\
U_{min}=0.6694297+0.2599712\;i 
\label{siegel}
\end{eqnarray}
We also find the modular transformed (see fig.2) minima under the action of 
the $Sp(4,Z)$ generator \footnote{The generators of $Sp(4,Z)$
are:$$\Biggl(\begin{array}{cc}
0&I_{2} \\
-I_2&0
\end{array}\Biggr)$$,$$\Biggl(\begin{array}{cc}
A& 0 \\
0&^{t} A^{-1}
\end{array}\Biggr)$$,$$\Biggl(\begin{array}{cc}
I_2 & B \\
0   & I_2
\end{array}\Biggr)$$
all $A\in GL(2,Z), B$ symmetric,integral}

$$\Bigl(\begin{array}{cc}
0&I_{2} \\
-I_2&0
\end{array}\Bigr)$$
which induces  
\begin{eqnarray}
T&\rightarrow& \frac{U}{TU-B^2}=\bar{U} \nonumber \\
B&\rightarrow& \frac{B}{TU-B^2}=\bar{B} \nonumber \\
U&\rightarrow& \frac{T}{TU-B^2}=\bar{T} \nonumber \\
\end{eqnarray}.

\begin{figure}
\epsfxsize=6.2in
\epsfysize=10.0in
\epsffile{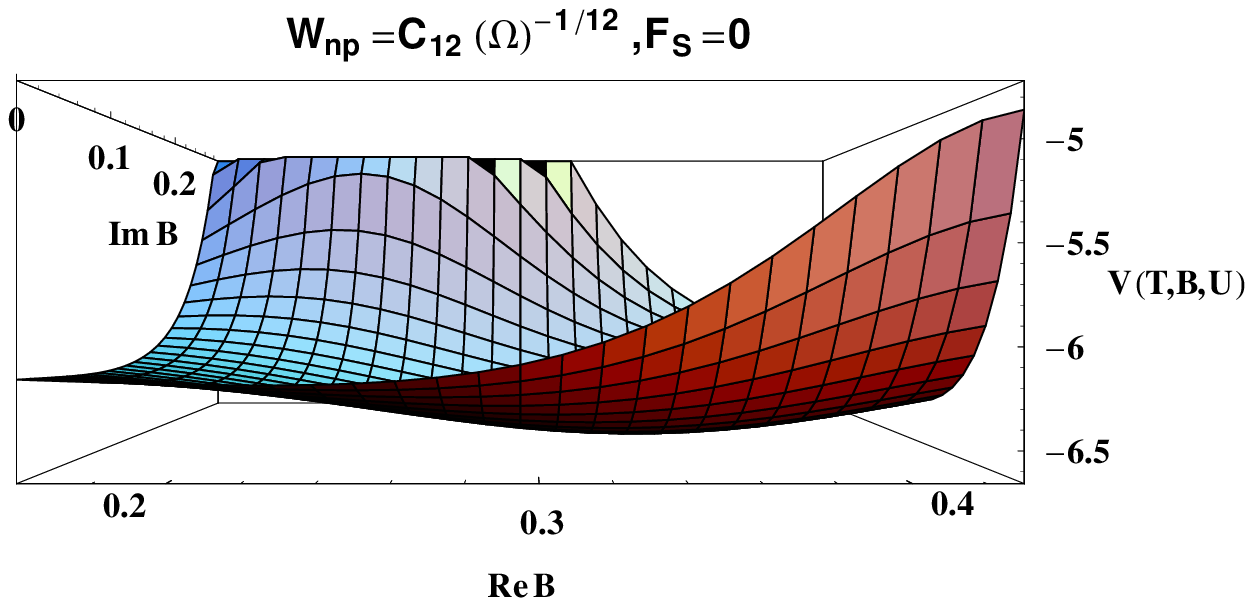}
\caption{Minimum of $V_{eff}$ in the Wilson line direction}
\end{figure}

\begin{figure}
\epsfxsize=6.2in
\epsfysize=9.0in
\epsffile{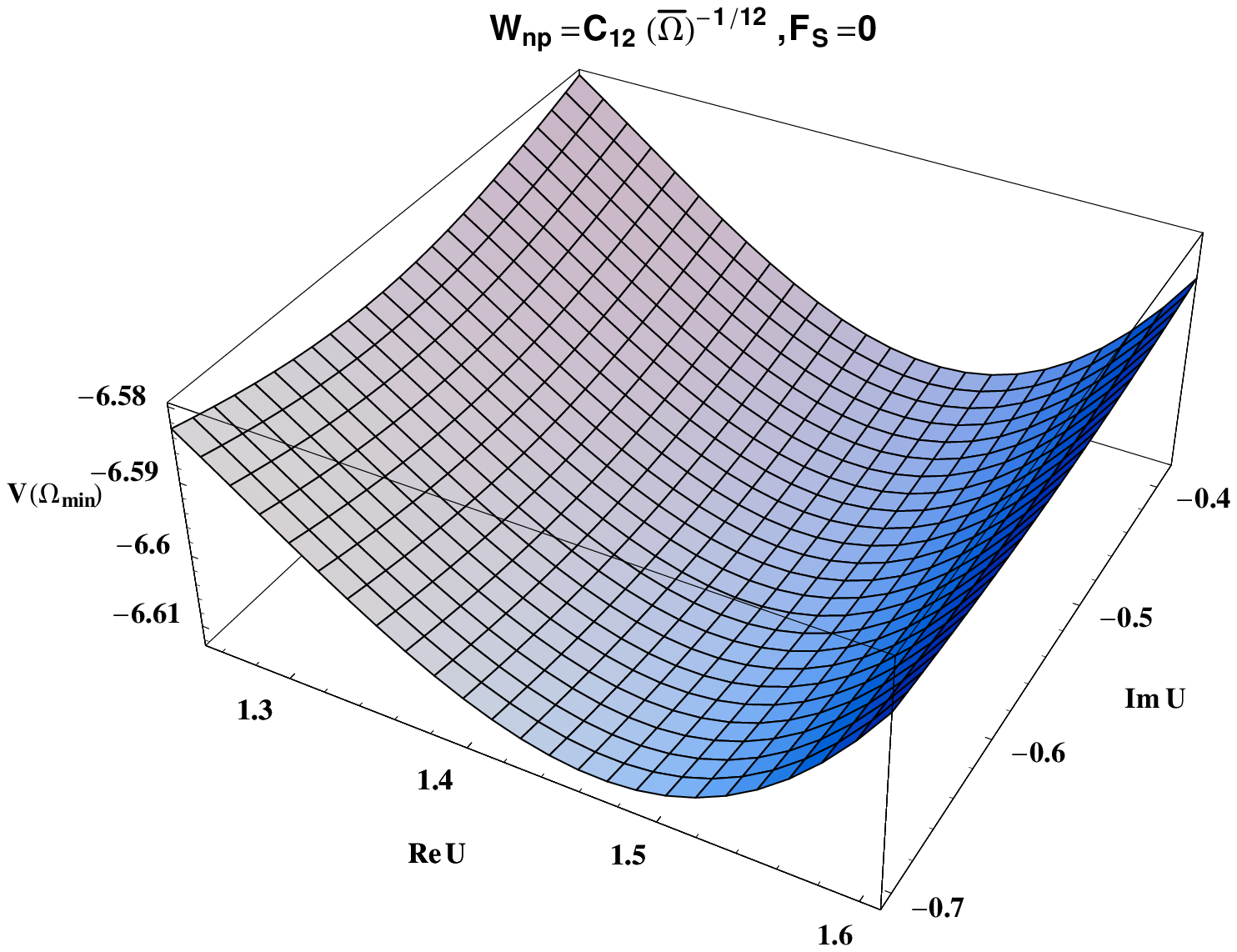}
\caption{Minimum of $V_{eff}$ in the $U$ direction, see Eq.(13)}
\end{figure}

One  might ask whether the minima obtained         
correspond to boundary or interior points of a particular 
``fundamental domain'' of $Sp(4,Z)$?
According to \cite{SIEGEL} the matrix of moduli $\Omega$ lies 
on the boundary of the generalized 
Siegel fundamental domain $\Im$ if,
\begin{equation}
\Bigl|{\rm det}(C\Omega+D)\Bigr|=1=\Bigl|
{\rm det}(-^{t} C \Omega^{'}+^t A)\Bigr|
\label{FD}
\end{equation}
for some choice of $$\Bigl(\begin{array}{cc}
A & B \\
C & D
\end{array}\Bigr)\in Sp(4,Z)$$ with $C \not =0$ and $\Omega^{'}$ 
the modular transformed matrix of moduli $\Omega$. 
The matrices $\Omega$ and $\Omega^{'}$ obtained from 
(12) and (13) respectively {\it do} satisfy these conditions and so 
are on the boundary. We regard this result as highly non-trivial.

%It is important to note that the above minima of $V_{eff}$ 
%are on the boundary of the generalized Siegel fundamental domain 
%$\Im$. It follows from Siegel's beautiful 
%theorem using the connection between the costruction of the fundamental 
%region of the elliptic modular group and the reduction theory of binary 
%quadratic forms \cite{SIEGEL}.
%This is a highly 
%non trivial result.

We also find the following (see fig.3-5) minimum
\begin{eqnarray}
T_{min}=1.29861+0.1191744 \;i \nonumber \\
B_{min}=1.09051+0.520288 \;i \nonumber \\
U_{min}=1.15447+0.194363 \;i \nonumber \\
\label{astir}
\end{eqnarray}
together of course with an infinite number of minima connected to 
it by Siegel modular transformations. 
For instance we find numerically the following minimum 
\begin{eqnarray}
T_{min}=0.56453+1.12787\;i \nonumber \\
B_{min}=0.243256+1.28333\;i \nonumber \\
U_{min}=0.721535+1.20548\;i \nonumber \\
\end{eqnarray}
generated from (\ref{astir}) by the symplectic 
Siegel modular transformations 
\begin{eqnarray}
T&\rightarrow& \frac{U}{TU-B^2} \nonumber \\
B&\rightarrow& \frac{B}{TU-B^2} \nonumber \\
U&\rightarrow& \frac{T}{TU-B^2} \nonumber \\
\end{eqnarray}
We also find the modular transformed minimum
under the transformation
\begin{equation}
\Omega \rightarrow A\Omega ^{t}A= \Omega^{'}=\Biggl(\begin{array}{cc}
U & B \\
B & T \end{array}\Biggr)
\end{equation}
with 
\begin{equation}
A=\Biggl(\begin{array}{cc}
0&1 \\
1&0\end{array}\Biggr)\in GL(2,Z)
\end{equation}

\begin{eqnarray}
T_{min}=U \nonumber \\
B_{min}=B \nonumber \\
U_{min}=T  \nonumber \\
\label{sastir}
\end{eqnarray}
The above minimum is an interior point since it does not satisfies the 
Siegel's equalities (\ref{FD}).

\begin{figure}
\epsfxsize=6.2in
\epsfysize=9.0in
\epsffile{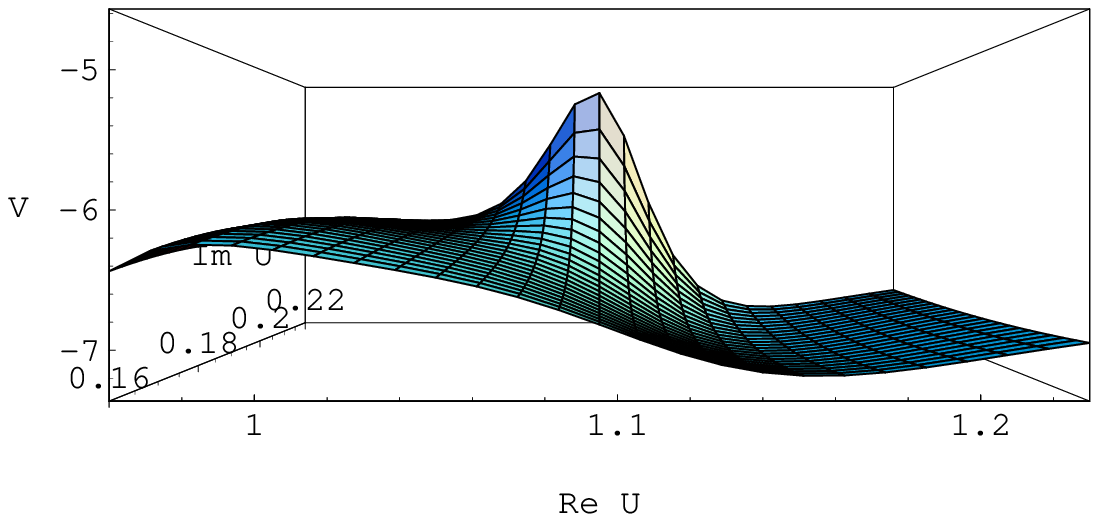}
\caption{Minimum of $V_{eff}$ in the $U$-direction, 
see Eq.(15) }
\end{figure}

\begin{figure}
\epsfxsize=6.2in
\epsfysize=9.0in
\epsffile{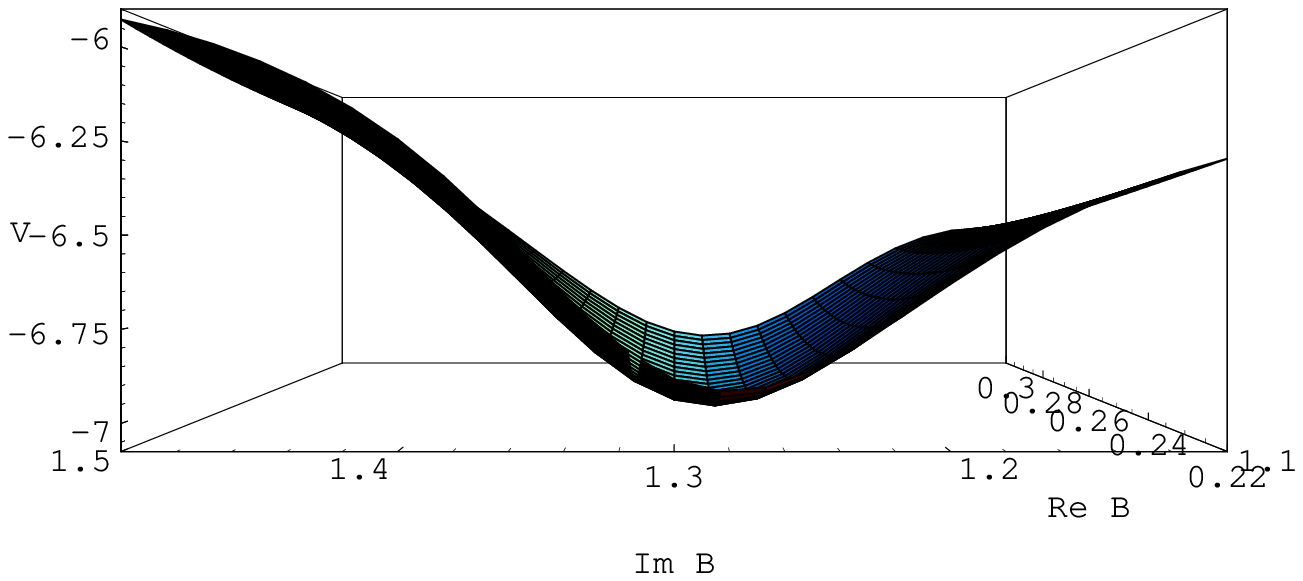}
\caption{Modular transformed minimum of $V_{eff}$ in the $B$-direction, 
see Eq.(16) }
\end{figure}

\begin{figure}
\epsfxsize=6.2in
\epsfysize=9.0in
\epsffile{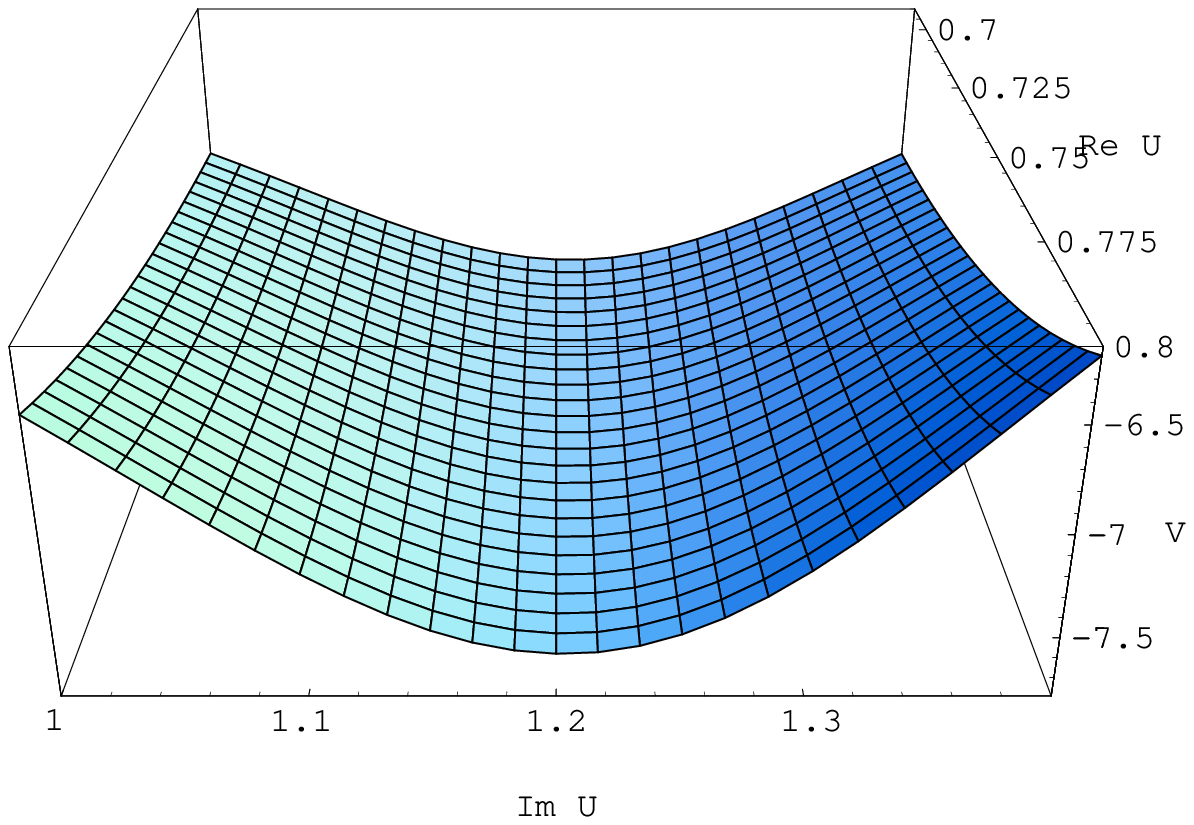}
\caption{Modular transformed minimum of $V_{eff}$ in the $U$-direction, 
see Eq.(16) }
\end{figure}

Interestingly, for large dilaton $F$-terms (i.e 
$|\frac{dP}{dy}|\gg 1$ and/or $0<\frac{d^2P}{dy^2}\ll 1$) we obtain 
familiar algebraic points of the $PSL(2,Z)$ and 
$\Gamma^{0}(3)$ modular groups.
For instance for $\frac{dP}{dy}=1.5$ and $ \frac{d^2P}{dy^2}=0.1$ we 
(see fig.6-7) obtain
\begin{eqnarray}
T_{min}&=&\sqrt{3} \nonumber \\
U_{min}&=&\frac{\sqrt{3}}{2}+\frac{1}{2} \;\;i
\end{eqnarray}
with the Wilson line 
\begin{equation}
B_{min}=\frac{\sqrt{3}}{2}+\frac{1}{2} \;\;i
\end{equation}
As  we shall see at this minimum $CP$-violation is zero
\footnote{This minimum also lies {\it on} the boundary of the 
generalized Siegel fundamental domain.}.
%Supersymmetric $CP$-problem is solved.

\begin{figure}
\epsfxsize=6.2in
\epsfysize=9.0in
\epsffile{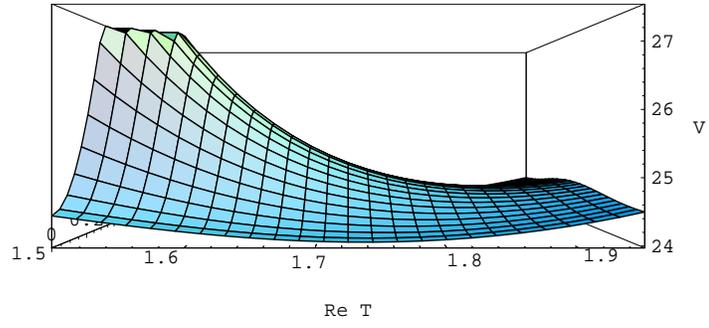}
\caption{Minimum of $V_{eff}$ in the $T$-direction at the familiar 
algebraic point of $\Gamma^{0}(3)$}
\end{figure}

\begin{figure}
\epsfxsize=6.2in
\epsfysize=9.0in
\epsffile{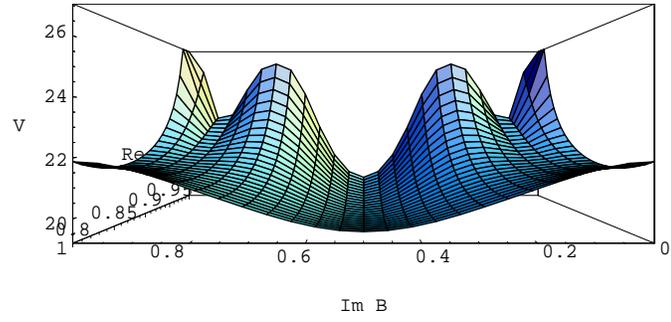}
\caption{Minimum of $V_{eff}$ in the $B$-direction at the 
familiar fixed point of $PSL(2,Z)$}
\end{figure}

We now calculate the $CP$-violation in the $A$ and $B$ terms. 
Unfortunately in this case (which corresponds to a case of an 
asymmetric orbifold 
\footnote{Although the form is known 
for asymmetric orbifolds in the 
absence of Wilson line moduli it is not known 
in their presence \cite{NAR}.}) the modular properties of 
the Yukawa couplings 
that appear in the trilinear soft $A$-terms 
in the presence 
of Wilson line moduli are unknown. In the absence of 
Wilson line moduli, when the modular groups 
are $PSL(2,Z)$ and $\Gamma^0(3)$ 
we could cast the twisted sector 
Yukawa couplings (calculated using 
conformal field theory techniques) in terms of Jacobi theta functions with 
definite modular properties \cite{Chun,BKL}.
Unfortunately we do not know how to generalize them to 
Siegel modular forms. However, we study the $CP$-violation arising 
from the $A$-terms when (i) the Yukawas $h_{\alpha\beta\gamma}$
have no modular dependence, and (ii) the Yukawas $h_{\alpha\beta\gamma}$
are proportional to the appropriate powers of ${\cal C}_{12}(\Omega)$. 
For $\mu_{W}$ we take the ansatz 
$\mu_W={{\cal C}_{12}(\Omega)}^{1/12}$ for the 
coupling $\mu_{W}\Phi_1 \Phi_2$  with 
$\Phi_1, \Phi_2$ both in the untwisted sector. Then in the limit 
$B\rightarrow 0$, $\mu_W\rightarrow 
\eta^{2}(T) \eta^{2}(U)$ 
consistent with earlier work by Antoniadis et al \cite{ANTO}.

Both $A$ and $B$-terms, in large regions of the parameter space 
with large auxiliary dilaton $F$-terms, 
lead to zero $CP$-violating phases. In this case the VEVs of 
the moduli fields including the Wilson line at the minimum of the 
effective potential are at familiar algebraic points of the 
$PSL(2,Z)$ and $\Gamma^{0}(3)$ modular groups. All of the soft terms 
as well as the $\mu$ term are {\it real}.
In the moduli dominated limit or in intermediate regions of the 
auxiliary dilaton-moduli field space soft $B$-terms lead to phases 
of order $10^{-2}-10^{-1}$. 
The properties of modular functions offer a pleasing explanation of the 
approximate $CP$-invariance of the soft supersymmetry-breaking terms.
In summary, the picture of the $CP$-structure of the soft supersymmetry 
breaking terms in the presence of a continuous Wilson line modulus is 
consistent with the picture that emerged from modular invariant 
effective 
actions in which only the metric moduli $T,U$ were present in the 
effective action \cite{BKL}. The resulting $CP$-phases are naturally small.

\section*{Acknowledgements}
This research is supported in part by PPARC.

\end{document}